\documentclass[10pt]{iopart}
\usepackage{iopams} 

\usepackage{graphics}
\usepackage{graphicx}

\begin{document}

\title{Liquid Scintillators for Large Area Tracking System}

\author{Yan  Benhammou$^{1}$, Erez  Etzion$^{1}$, Gilad  Mizrachi$^{1,2}$, Meny  Raviv Moshe$^{1}$ and Yiftah  Silver$^{2}$}

\address{$^{1}$ Raymond and Beverly Sackler School of Physics \& Astronomy, Tel Aviv University, Tel Aviv 69978, Israel,\\ $^{2}$ Rafael, Advanced Defense systems LTD, Haifa, Israel}

\ead{erez@cern.ch}

\vspace{10pt}
\begin{indented}
\item[]January 2020
\end{indented}

\begin{abstract}

We report on studies of non-toxic scintillating liquid useful for large surface  detectors.   Arrays of liquid scintillators  offer  a rather simple  tool for detecting charged particles traversing a surface and tracking their path through a defined volume. 
Insertion of wavelength shifting fibres along the liquid scintillating containers  significantly improves the light collection at the two ends of the scintillators.
We have demonstrated that we can achieve timing resolution of $\mathcal{O}(1$~ns) allowing  good spatial resolution. 
Liquid scintillators with fibres read by Photo-multipliers or SiPMs provide  an  inexpensive alternative technology which  suits well the requirement of the MATHUSLA experiment tracking system. 
\end{abstract}

%
\vspace{2pc}
\noindent{\it Keywords}: High Energy Physics, Particle Detectors, Liquid Scintillators

\submitto{Journal of Physics Communication}


\ioptwocol
\section{Introduction}
\label{sec:Introduction}

The discovery of the Higgs boson at the {L}arge {H}adron {C}ollider (LHC)~\cite{Aad:2012tfa, Chatrchyan:2012xdj} marked a major success to the ATLAS and the CMS detectors, and  opened a new direction of searches  
for extensions to the Standard Model (SM) of particle Physics. 
The attention was turned to Beyond the SM (BSM)  phenomena not addressed by the SM. Many theoretical frameworks that shed more light on the fundamental mysteries of the SM are expected to include new particles discoverable at the LHC. 
However, to date all direct BSM searches have produced null results.  

\begin{figure}[h!]
\begin{center}
\includegraphics[width=0.47\textwidth]{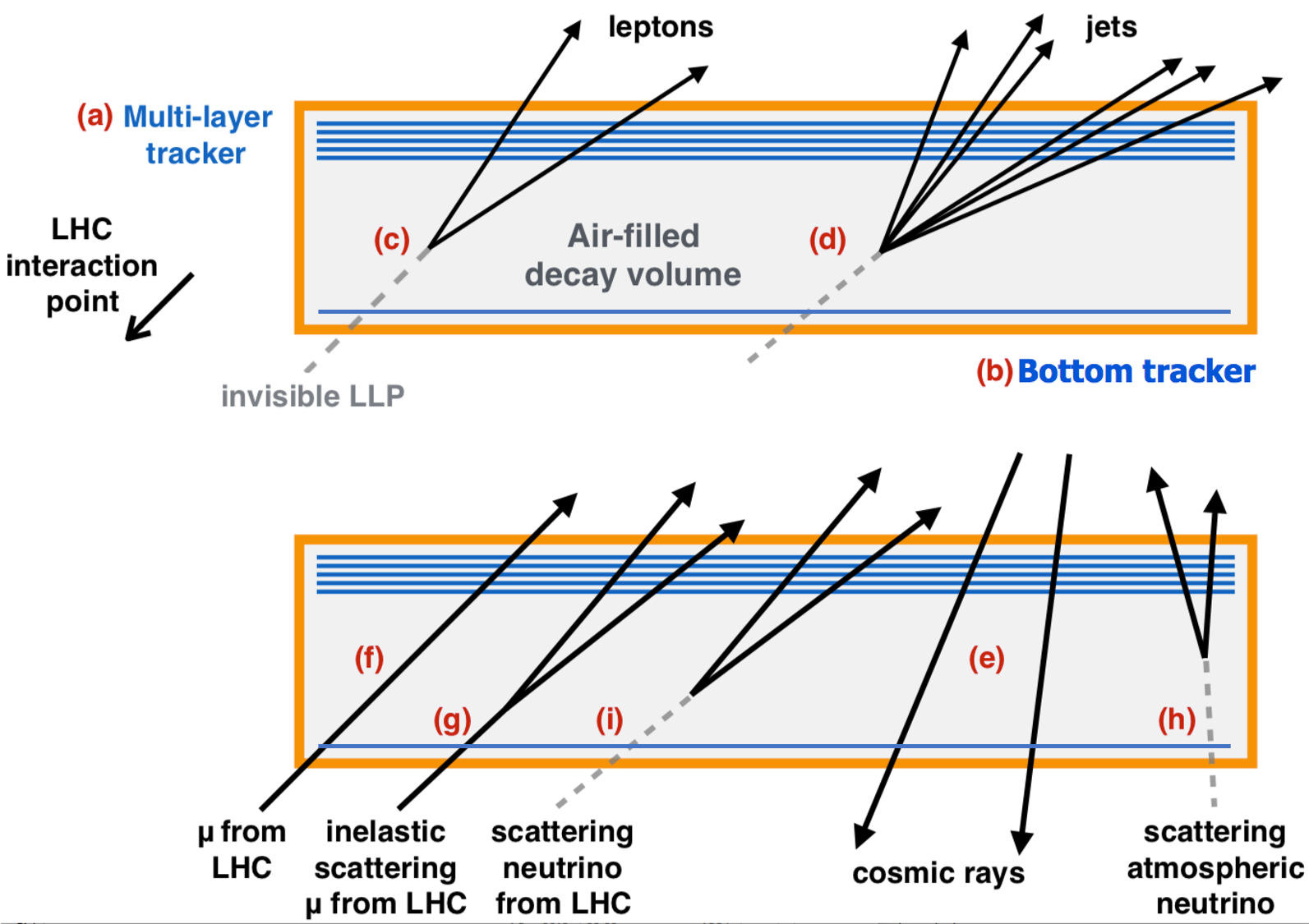}
\end{center}
\caption{ Schematic illustration of LLP decay signal (top) and main backgrounds (bottom) in a MATHUSLA-like detector consisting of an upper tracker (a) above an air-filled decay volume and bottom tracker (b) on the floor. (c) and (d) are illustrations of upwards going signals, where (e)-(i) depict potential background signatures.}
\label{f.bgsummary}
\end{figure}

A reason for the  lack of signs for new physics could be that the  searches  
have been focused on the production of particles, which promptly decay 
into visible final states very close to the interaction point. 
A less explored option is that BSM signals first appear as new {L}ong-{L}ived {P}articles (LLPs) that decay into SM final states some macroscopic distance away from the production point. 
LLPs, which arise due to some mechanism suppressing the decay rate of the particle, may be generic extensions to the SM.
LLPs are a favorable prediction of a wide range of theories, such as  hidden sectors or hidden valleys~\cite{Strassler:2006im}, Neutral Naturalness~\cite{Chacko:2005pe} and Super Symmetry~\cite{PhysRevD.40.2987, Dreiner:1997uz}, models of Dark Matter e.g.~\cite{Cui:2012jh} and neutrino extensions such as ~\cite{Minkowski:1977sc,Mohapatra:1979ia}, all of which, can address  the  fundamental shortcomings of the SM. 
Neutral LLPs with average decay length above $\approx$100~m are particularly difficult to probe, as the sensitivity of the LHC main detectors is limited by challenging backgrounds, triggers, and small acceptances. MATHUSLA is a proposed project for a minimally instrumented, large-volume surface detector above the  LHC main detectors ATLAS or CMS. It would search for neutral LLPs produced in High Luminosity LHC collisions by reconstructing displaced vertices in a low-background environment, extending the sensitivity of the main detectors by orders of magnitude in the long-lifetime regime. Neutral LLPs produced at the LHC interaction point travel upwards and decay inside the air-filled\footnote{This is required to allow for the detection of LLPs that decay hadronically.} decay volume, giving rise to displaced vertices of upwards-traveling charged tracks that are reconstructed by a highly-robust multi-layer tracking system near the roof. The basic principle of the MATHUSLA detector concept is illustrated in figure~\ref{f.bgsummary}. The Physics case was very well summarised and discussed in details at the white paper~\cite{Curtin:2018mvb}. 
 
 The original simplified design proposed in~\cite{Chou:2016lxi}  assumed five layers of Resistive Plate Chambers (RPCs), which are suitable for economical coverage of  very large areas ($200 \times 200$~m$^2$) with good position and timing resolution. In this work we assess the possibility of using  liquid scintillator for the active material in the tracker layers.  Another promising technology  being considered in the MATHUSLA Letter of Intent~\cite{Alpigiani:2018fgd} and the project proposed to the European strategy group~\cite{Lubatti:2019vkf}  is plastic  scintillators. Here we look at Liquid scintillator detectors which may serve as another potentially promising alternative to gaseous detectors. This is due to their relative low price and the relative ease in which large volume detector can be constructed.

.

\section{Methodology}
\label{sec:Measurements}

\subsection*{Liquid Scintillators}
\label{sec:LiquidScintillators}
Liquid scintillators are made of solutions (homogeneous mixture) of scintillating solvent with additional materials dissolved in it, in order to shift the output emission into visible light.
Traditionally, liquid scintillators are made with hazardous materials such as toluene, xylene and benzene as the organic solvent (see e.g.~\cite{doi:10.1139/p58-006} or compilation of several options in~\cite{Barton_1962}), this makes the preparation and handling of detectors difficult. We took a different approach of using Linear Alkyl Benzene (LAB),  the dominant precursor of biodegradable detergents~\cite{Kosswig:IndustrialChemistry}, which is much cheaper, easy to get and less hazardous than commonly used liquid scintillating materials. The clear advantage is that it makes safe handling easy. In addition, LAB has a very good optical transparency ($\approx 20$~m) and low amount of radioactive impurities \cite{CHEN:200565}. The typical solutes are fluors such as {\it 2,5-Diphenyloxazole} (PPO) with output spectrum peak at 385~nm, and wavelength shifters such as {\it 1,4-bis(5-phenyloxazol-2-yl) benzene} (POPOP) or {\it 1,4-bis (2-methylstyryl) benzene} (MSB), these are added to the organic scintillators to shift their emission spectrum into the visible range, with emission maximum at 410~nm (POPOP) and 420~nm (MSB). 
In order to test the feasibility and appropriateness of liquid scintillators for a large area high resolution charged particle detectors, we have conducted several experiments using two different geometries. The geometries used are plane and tube containers both filled with different compositions of liquid scintillating materials. 

\subsection{Liquid scintillator plane}
\label{ss.LiquidpPane}
\begin{figure}[!htbp]
\begin{center}
\includegraphics[width=0.47\textwidth]{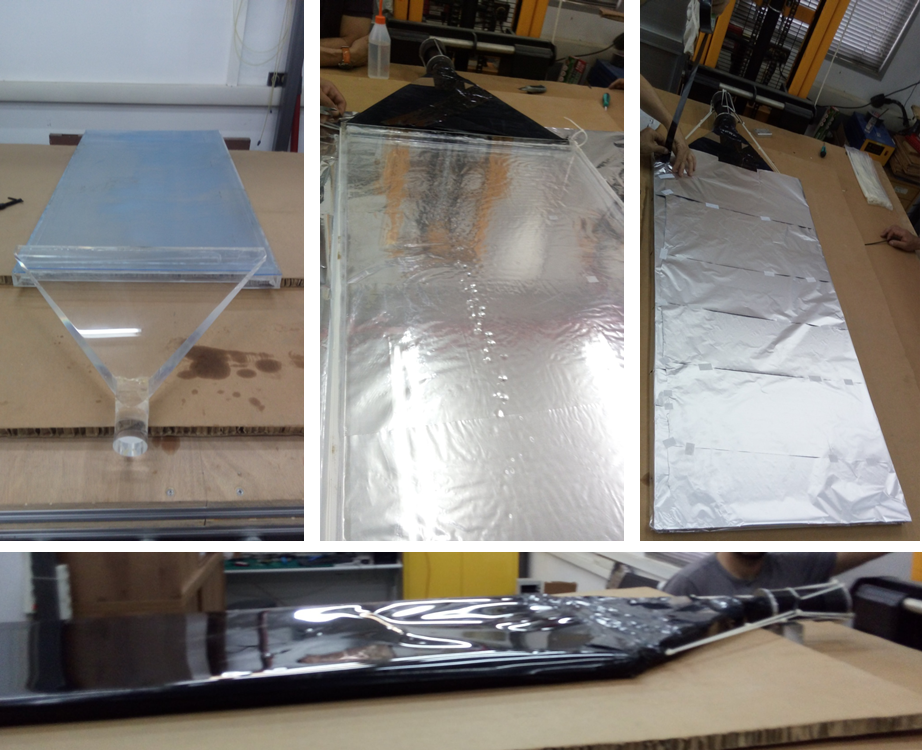}
\end{center}
\caption{
Pictures of the liquid scintillator plate and the triangle light guide during its production process: top left - the perspex container filled with the mixed liquid, top right - wrapped with aluminium foil and down wrapped with light tight black cover.}
\label{f.LiquiPlatePhotos}
\end{figure}

\begin{figure}[!htbp]
\begin{center}
\includegraphics[width=0.47\textwidth]{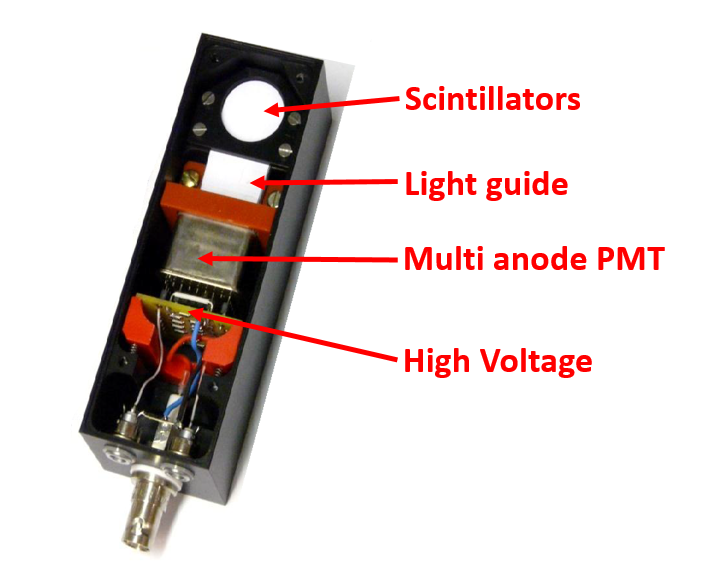}
\end{center}
\caption{
The hodoscope used for the efficiency measurements (top cover removed), the main components are described.}
\label{f.HodoscopePhotos}
\end{figure}

A $60\times 120$~cm$^2$ plane,  1.2~cm deep of liquid scintillator (figure~\ref{f.LiquiPlatePhotos}) was constructed using a perspex box covered with aluminium foil. The light emitted from the liquid scintillator is tunnelled by a triangular light guide into a $2"$ Hamamatsu R329-02 PMT\footnote{https://www.hamamatsu.com/eu/en/product/optical-sensors/pmt/index.html} which is sensitive to light from 300~nm to 650~nm peaking at 420~nm. The liquid scintillator plate efficiency to cosmic ray muons was measured with a dedicated hodoscope. The hodoscope includes two 3~mm thick round NaI(TI) scintillator counters (diameter 12.5~mm one on top of the other with a gap of 5~mm between the two) readout by multi-anode photomultiplier with a gain of $10^6$. A picture of the hodoscope is shown in figure~\ref{f.HodoscopePhotos}. Signals from both hodoscope scintillators are discretized and their coincidence serves as a trigger. The efficiency ($\varepsilon$) is then estimated by  using equation \ref{eq.Efficiency}, which is the number of signals collected in the liquid scintillator plane in a 100~ns time window  following the hodoscope trigger ($N_{\mu}$) divided by the total number of triggered events ($N_{t}$). 

\begin{equation}
\label{eq.Efficiency}
\varepsilon  = 100 \times \frac{N_{\mu}}{N_{t} }~\%
\end{equation}

\subsection{Liquid scintillator tubes}

\begin{figure}[!htbp]
\begin{center}
\includegraphics[width=0.47\textwidth]{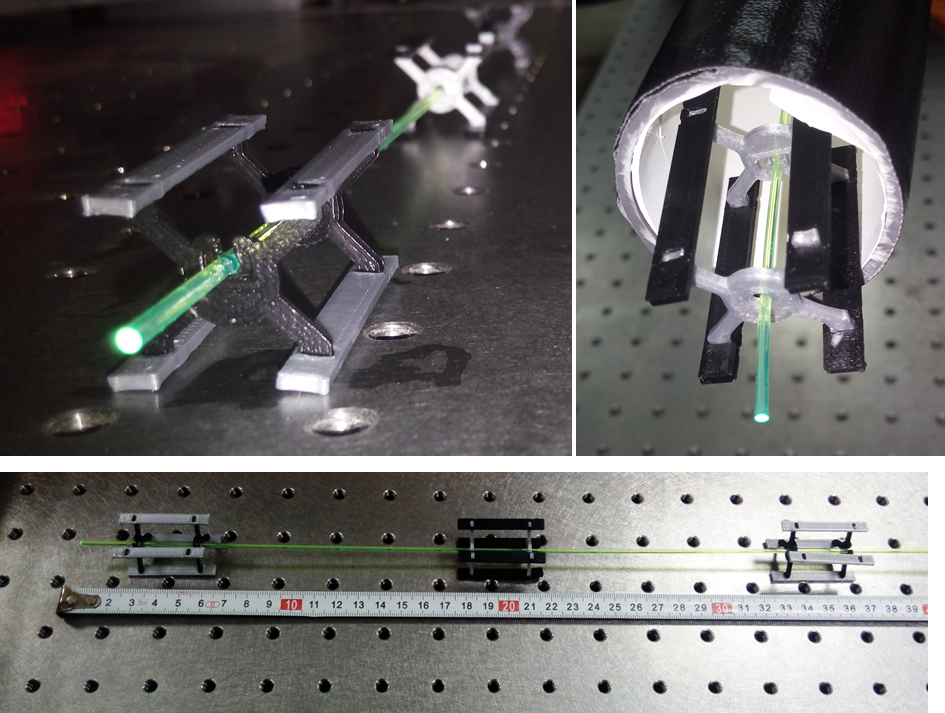}
\end{center}
\caption{
Pictures of the WLS fibre used in the tube setup efficiency measurements. The fibre is placed in the specially designed holders.}
\label{f.FiberPhotos}
\end{figure}

\begin{figure}[!htbp]
\begin{center}
\includegraphics[width=0.48\textwidth]{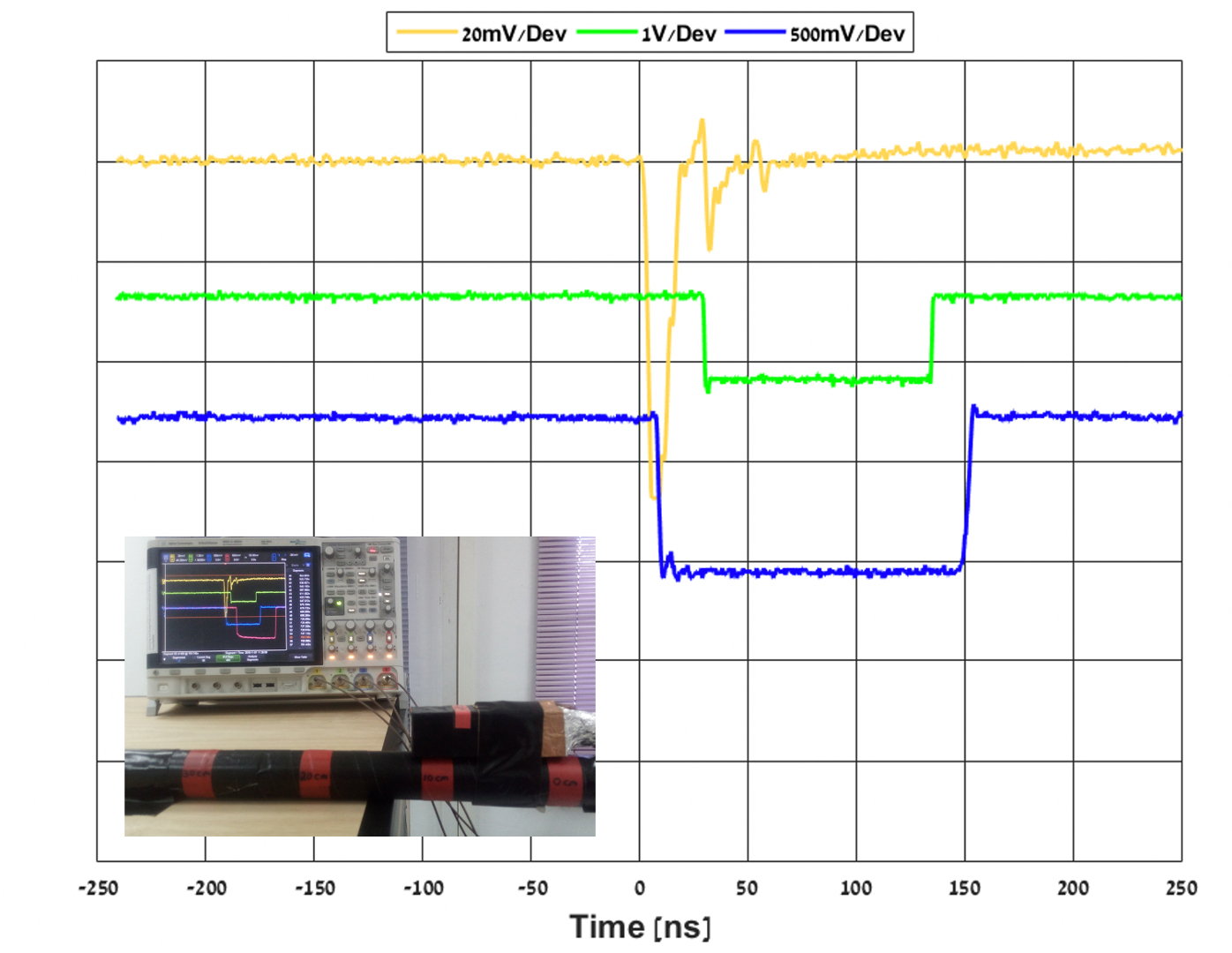} 
\end{center}
\caption{A scope display screenshot of typical signal during a liquid-scintillator-tube relative efficiency measurement. 
 The yellow (blue) trace is the signal (discritized signal) of the liquid scintillator as read by the PMT, while the green trace is the hodoscope trigger. The inset depicts the setup of this measurement, where the hodoscope is placed above the tube 10~cm away from the PMT. The additional magenta line in the picture is the coincidence between the hodoscope trigger and the liquid scintillator signal.}
\label{f.LiquidTubePhotos}
\end{figure}

In order to examine the behaviour of the liquid scintillator in tube geometry we have conducted a set of measurements using  1~mm Y11(200) blue to green Kuraray Wave Length Shifter (WLS) fibre with attenuation length  $\lambda > 3.5$~m  ($\lambda$ given by ref~\cite{Kuraray} fitting exponential function to the light collected varying the fiber length, $x$, as $I(x)=I(0)exp(-\frac{x}{\lambda})$). The fiber runs inside the liquid scintillator placed in a box shape 5~cm $\times$ 10~cm, 1~m long aluminum tube. Dedicated holders, shown in figure~\ref{f.FiberPhotos} keep the fibres centred as well as coupled to the PMT positioned in the end of the tube.
Utilizing the same hodoscope as in section~\ref{ss.LiquidpPane} above the tube (as depicted in the inset of figure~\ref{f.LiquidTubePhotos}) we estimated the efficiency of reading signals generated by cosmic ray muons. Figure~\ref{f.LiquidTubePhotos} depicts a scope screenshot of a typical signal read with this setup. Equation~\ref{eq.Efficiency} is not sufficient for estimating the absolute efficiency of the tube shape scintillators since the angular acceptance of the hodoscope is larger than the  tube itself  in a non trivial way. As a consequence, relative efficiency was estimated with respect to the efficiency at the nearest position to the PMT (set as $100\%$). The relative efficiency graph obtained for different scenarios ( i.e. different concentrations of MSB and distances from the PMT) is shown in figure~\ref{f.LiquidTubeEfficiency}.

\section{Results and Discussion}
\label{sec:Results}
\begin{figure}[!htbp]
\begin{center}
\includegraphics[width=0.47\textwidth]{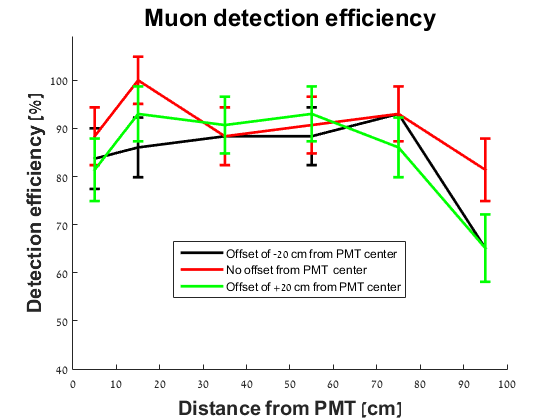}
\includegraphics[width=0.47\textwidth]{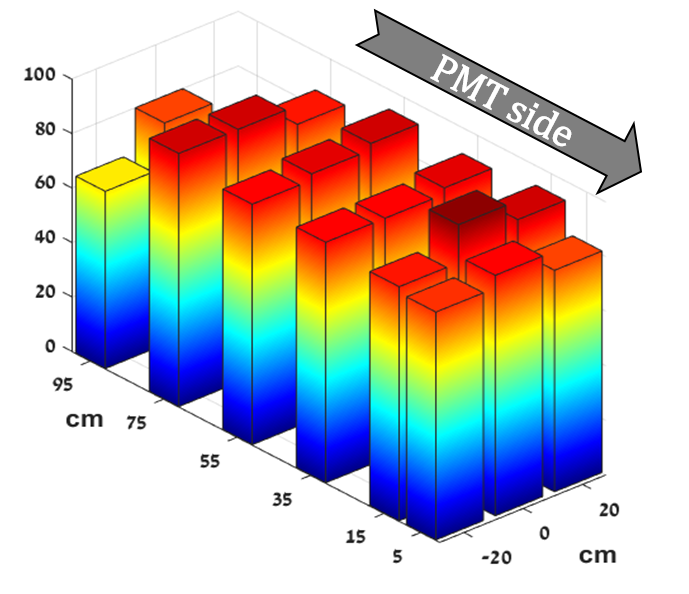}
\end{center}
\caption{
top: The detection efficiency as as a function of the distance from the PMT scanning on three  lines along the scintillator: the red line in the center of the scintillator while the black and red on parallel lines left or right of it and 20 cm away from the center line.

bottom: A 2D muon detection efficiency map obtained with a liquid scintillation plane. Each data point represents a different hodoscope location on top of the plane. Distance from the light guide and the PMT directions are indicated.}
\label{f.LiquidPlateEfficiency}
\end{figure}

We took data generated by cosmic ray muons while scanning the liquid scintillating plane by shifting  the hodoscope to different locations above the surface. The result are the light collection efficiency at various positions with respect to the PMT depicted at figure~\ref{f.LiquidPlateEfficiency} (top) as well as in a  2D efficiency map shown in figure~\ref{f.LiquidPlateEfficiency} (bottom). With this setup we managed to obtain a rather homogeneous and high  efficiency rate of signal detection with an efficiency drop at the edges of the scintillator. The axes in the two figures indicate the distance from the middle front face of scintillator plane.    
As seen the efficiency is between $85\%$ and $95\%$ everywhere except for the back corners of the box. The somewhat poor light collection from the back side (furthest from the PMT) of the plane is due to lower level of liquid scintillator material (caused by a slight tilt of the experimental setup). The poor light collection from the rear part manifests as efficiency at the level of $65\%$ to $85\%$.

\begin{figure}[!htbp]
\begin{center}
\includegraphics[width=0.47\textwidth]{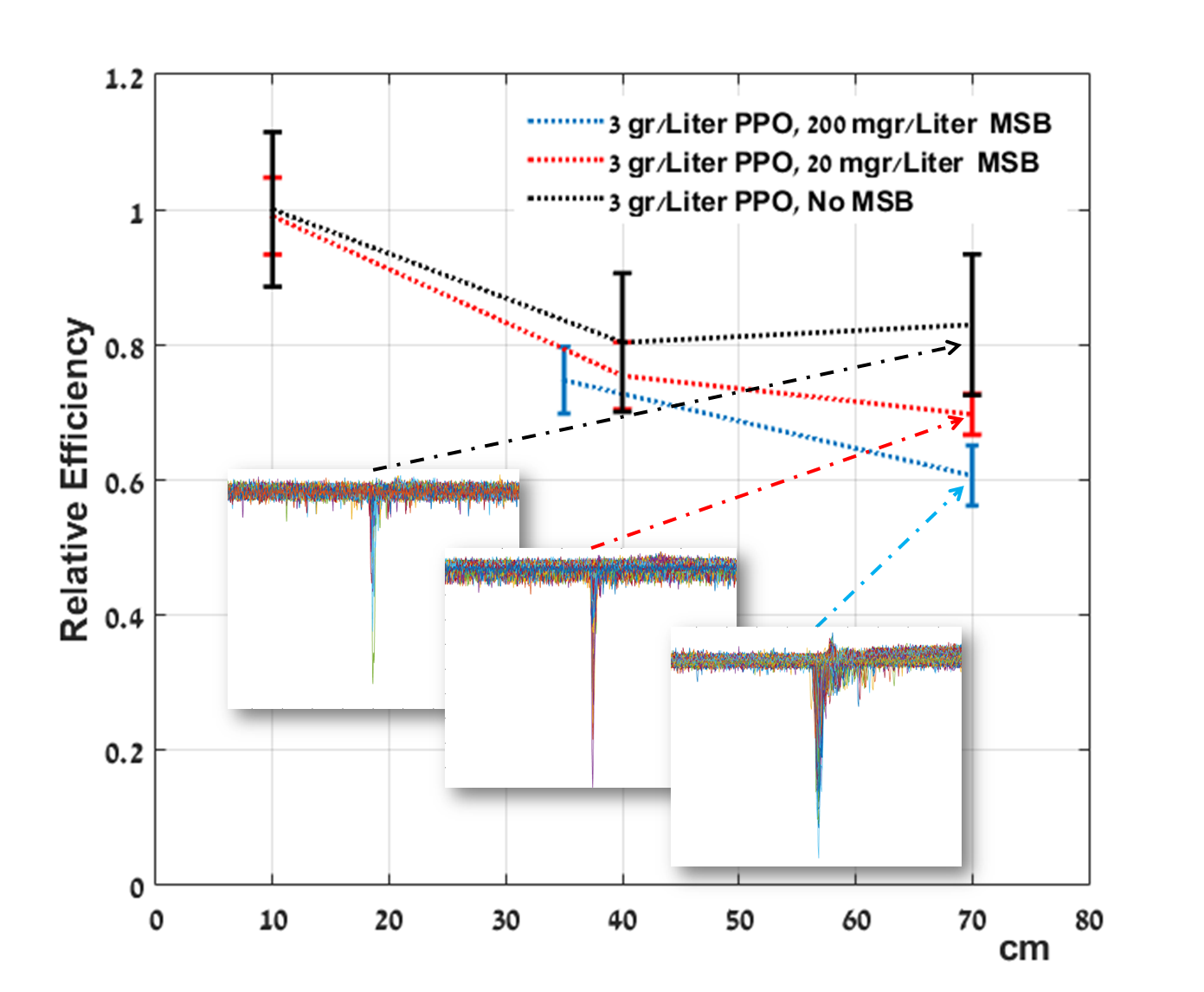}
\end{center}
\caption{Relative muon detection efficiency obtained in the tube geometry with different concentrations of MSB. The x axis indicates the hodoscope position with respect to the  PMT (in cm). The three insets in the figure show an overlay of all the signals obtained 70~cm away from the PMT.
}
\label{f.LiquidTubeEfficiency}
\end{figure}

Figure~\ref{f.LiquidTubeEfficiency} shows the relative muon detection efficiency obtained in the tube geometry with different concentrations of MSB in various distances from the PMT. All measurements are normalized to the maximal efficiency, obtained when the hodoscope was placed 10~cm away from the PMT (set to $100\%$). The three insets in the figure show an overlay of all the signals obtained 70~cm away from the PMT. The signals obtained in this setup (from the liquid scintillation tube) are sufficiently fast with full width at half maximum (FWHM) of $\mathcal{O}(10$~ns) but the rise time (the time required for the signal to increase from 10\% to 90\% of the steady state value) is of $\mathcal{O}(1$~ns).

While varying the distance between the triggered events to the PMT in our setup and changing the concentration of the MSB in the liquid composition of the tube we observed two effects:

\begin{enumerate}
    \item The efficiency drops with the distance from the PMT. 
    \item The efficiency drops as the concentration of the MSB increases (with constant PPO concentration). 
\end{enumerate}

Hence, we deduce the following: first, the WLS fibre coupling to the PMT in our setup might not be optimal, second, the addition of MSB is not necessary in order to achieve high detection efficiency. The reason why the addition of MSB degrades the efficiency could be  related to the spectral response of the PMT. It  is dominated in the  violet to ultra violet range, already produced by the PPO ($385$~nm). The additional MSB absorbs photons emitted by the PPO and shifts it to $420$~nm with lower efficiency. This process lower the overall efficiency (see ref~\cite{PMTHandbook}).
In parallel to the studies of liquid scintillators we have examined in the laboratory the usage of extruded scintillating bars read with  Silicon  Photomultipliers (SiPMs), similar to the MINOS experiment setup~\cite{Evans:2013pka}. Like in the long tubes of liquid scintillators setup, WLS  fibres are used to increase the efficient length range of the bars. Preliminary studies show that one can achieve  efficient light collection with fibres as long as three to five meters. The results of these studies will be soon published in a more detailed paper.

\section{Summary}
\label{sec:Summary}

The MATHUSLA experiment is a dedicated large-volume displaced vertex detector for the high luminosity LHC runs, designed to be positioned on the surface above  one of the main LHC experiments.
It is aimed at searching for long-lived particles escaping detection of the underground experiments  with up to several orders of magnitude better sensitivity than the current experiments.
The distance from interaction point and a desirable angular coverage determines a large detection area of a $\mathcal{O}(10^4$~m$^2$). In our studies we have seen that a large area of tracking system can be realised by layers made with non-toxic liquid scintillators (PPO and MSB dissolved in LAB).  A clear advantage of this solution is the simplicity of the construction and detector maintenance and the low cost of the materials. The studies showed that this setup requires further optimization of the exact liquid components. While FWHM of the detected signals is of $\mathcal{O}(10$~ns) the fast measured rise time allowing good spatial and temporal resolution. Light collection efficiency for large distances can be improved with the insertion of WLS fibers to liquid scintillators containers. The combination of a few meters long liquid scintillators tubes with WLS fibers, read by SIPMs, provides a inexpensive good alternative building block for the MATHUSLA experiment.

\section*{Acknowledgment}

The authors would like to thank the PAZY foundation.

\section*{References}
\bibliographystyle{iopart-num}
\bibliography{Mathusla-Liquid-Scintillator.bib}
\end{document}